\documentclass[a4paper,10pt]{article}
\usepackage[utf8]{inputenc}
\usepackage{graphicx}
\usepackage{subcaption}
\usepackage[english]{babel}
\usepackage {natbib}
\usepackage{array}
\usepackage{lscape}
\usepackage{float}
\usepackage{xcolor}
\usepackage{booktabs}
\usepackage[left=3.75cm, right=3.75cm, top=2cm, bottom=2cm]{geometry}
\usepackage[super]{nth}
\usepackage{fourier} 
\usepackage{makecell}
\usepackage{keyval}
\usepackage{amsmath}
\usepackage{fixltx2e}
\usepackage{longtable}
\usepackage{authblk}
\title{On the heterogeneity of urban expansion profiles in Europe}
\author[1]{Paul Kilgarriff}
\author[2]{Rémi Lemoy}
\author[1,3]{Geoffrey Caruso}
\affil[1]{Luxembourg Institute of Socio-Economic Research (LISER), Luxembourg}
\affil[2]{University of Rouen, IDEES Laboratory UMR 6266 CNRS, France}
\affil[3]{University of Luxembourg}

\date{Version:  \today}

\begin{document}

\maketitle

\begin{abstract}
The difference of a city's artificial land use (ALU) radial profile to the average ALU profile is examined for 585 European cities. Using Urban Atlas 2012 data, a radial (or monocentric) approach is used to calculate a city's land use profile in relation to distance to the city centre. A scaling law is used which controls for city size and population. As a consequence, cities of varying degrees of size can be contrasted in a comparable way. Utilising the mean ALU profile for the entire sample of 585 cities, the difference to the mean profile is calculated for each city. Using these differences allows us to examine heterogeneity of the ALU across European cities but also examine these differences within cities. We utilise city groupings by city size and country to attempt to understand these differences. Combining Urban Atlas and Corine Land Cover data, the impact of water on the ALU profiles is examined. A city classification is also introduced which considers the difference to the average curve. Ordering methods are used to visualise cities within these classifications. Results highlight the level of heterogeneity between cities. Removing water, we can see that the cities with the highest levels of water have a higher level of ALU on average. Spain and France are found to have contrasting levels of ALU, Spanish cities having below average ALU and France above average. Using seriation techniques enables us to group and order cities into a typology which can be used to benchmark cities. 
\end{abstract}

\section{Introduction}

There is a romantic notion that a single entity of a European city exists, one which has a monocentric structure alongside high density and compactness. Such a narrative could be considered naive as it assumes European cities are homogeneous. Although, structures exist in European cities which set them apart from US cities \citep{schneider2008compact,patacchini2009urban}. Within Europe, many cities possess a common urban form, with a similar gradient of built-up land in relation to distance to the centre, irrespective of their size \citep{Lemoy2018}. However, high levels of heterogeneity between cities remain. The challenge is how this heterogeneity can be measured. Two types of heterogeneity exist in relation to cities, inter (between cities) and intra (within cities between areas). Given the overall monocentric nature of European cities \citep{oueslati2015determinants}, urbanisation forms a radial structure, and there is a decrease of density as one moves away from the centre, as theorised by urban economics \citep{Alonso1964new,Mills1972new,Muth1969new,fujita1989urban1}. The aim of this paper is to describe this decreasing density with distance for European cities, and how this differs from the overall average urban density gradient. The work of \citet{Lemoy2018} is updated for 2012 to examine how 585 European cities are different from the 'average' European city. This examination of heterogeneity is important as it has consequences in relation to sustainability and efficient allocation of resources.

Examining differences between a city's artificial land use gradient and the average gradient, will identify distances where a city is more or less urbanised. It matters whether a city is more urbanised near the core or periphery, particularly in relation to aspects such as sustainability and accessibility. The accessibility aspect of a city is important, and mainly related to the distance to the city centre and how compact a city is at various distances. The level of compactness will have positive and negative influences on the sustainability of cities \citep{lin2006does,neuman2005compact,burton2003compact}. 
Densification of urban areas is associated with overcrowding and air pollution, while reversing patterns of decentralisation towards a compact city may not be feasible in all situations \citep{breheny1995compact}. Despite this, there are obvious advantages and disadvantages of higher urban density related to mobility, resources, social equity, economic output and energy consumption (see \citet{boyko2011clarifying} for a comprehensive review). The more compact a city is, the greater the level of economy of scale efficiencies, especially for public transport provision. A trade-off exists between capacity, frequency and location. Modes such as train or light rail have a higher capacity compared to buses, however costs are also higher therefore limiting the number of stations and stops. The cost effectiveness of different modes of shared transport is therefore dependant on the compactness of a city. A compact city also encourages walking and cycling for shorter distances. A city may be expanding in ways which make it difficult for policy makers to provide adequate services on limited budgets. In that way, heterogeneity of cities will impact on the sustainability levels if there are inefficient resource allocation and losses as a result of being over- or under-urbanised.

This is not the first time that urban forms and heterogeneity have been examined in a European context. Interestingly some earlier studies were also monocentric or radial, focusing on the internal structure of cities \citep{berry1963urban, Clark1951, fooks1946x}. Radial analysis can provide rich data on the level of urbanisation within a city. Recent studies however ignored the distance effect and focused on landscape metrics, with a few exceptions \citep{jiao2015urban, guerois2008built}. The rise in the popularity of using landscape metrics was largely driven by the focus on urban sprawl and the organisation of land \citep{eaa2006urban,eea2016urban}, fragmentation indices were viewed as the best method of examining both the level of urbanisation but also how fragmented parcels of land were. Measures used in ecology such as the Shannon index \citep{shannon1948mathematical} and Simpson index \citep{simpson1949measurement} offer a means of measuring fragmentation quantitatively. Landscape metrics place a greater emphasis on the local context as opposed to the overall land use profile. Whereas in ecology the local environment is important for biological diversity, richness and regularity \citep{tucker2017guide}, in a human context, movement in a city and accessibility to the centre is crucial \citep{Rode2017}. Although landscape metrics are useful, the lack of distance to the centre makes them less informative. It is not only the 'what' but also the 'where' which is important in relation to urbanisation and compactness of cities. Accounting for both the level of compactness and where this compactness occurs, is required to access the overall sustainability of cities. Landscape metrics will vary both within and between areas in a city and its surrounding areas. The use of landscape metrics to measure urban sprawl and compactness should act as a complement to the distance/radial based approach \citep{irwin2007evolution}. 

Previous studies have examined the level and (temporal) change in built-up area, urban land or residential area. Others have examined urban areas using landscape metrics or using an urban density, population change, urban expansion, urban sprawl and soil sealing approach. \citet{patacchini2009urban} used data from the Urban Audit (Eurostat) to examine the characteristics of urban sprawl using measures of land area, population density and the ratio of working age population to land area for 263 European cities over the period 1991-2001. Others have used landscape metrics to examine the impact of local culture \citep{netto2020form} on urban form. \citet{kasanko2006european} used a sample of 15 cities to examine urban expansion and heterogeneity between the 1950s to 1990s using the MOLAND model (JRC) \citep{engelen2004moland}. They found the surroundings and typography of a city and the historical onset of urbanisation are some of the determinants of city heterogeneity. \citet{eaa2006urban} also using data from the MOLAND model (JRC), examined sprawl at the UMZ level but introduced an additional distance measure in the form of three buffer zones outside of the UMZ (0-5km, 5-10 km, 10-20km). \citet{schwarz2010urban} used data from CORINE and the Urban Audit to measure urban form using several landscape and socio-economic metrics for 231 European cities. Utilising cluster analysis, statistically significant differences were found with respect to urban form among cities. 

\citet{siedentop2012sprawls} measured urbanisation within 20km grid cells using CORINE data to compare urban land use change over time both within and between 26 European countries. Heterogeneity within and between countries was found. \citet{turok2007trajectories} showed the majority of European cities experienced growth in population (1960-2005), with regional differences experienced. \citet{angel2011dimensions} used MODIS 500m land cover data along with UN and \citet{brinkhoff2010city} population data to examine urban land cover and population for 3,646 world cities using a morphological definition of cities. \citet{wolff2018compact} delineated cities using a combination of Urban Morphological Zone (UMZ) and density thresholds to examine the relationship between residential area change and population change in 5,692 European urban areas. \citet{oueslati2015determinants} examined the determinants of urban sprawl for 282 European cities at the LUZ level, for the years 1990, 2000 and 2006. Two indices of urban sprawl where used, one measure reflecting the scale of the change (growth of artificial area) and the other fragmentation (scatter index). Although the findings are useful particularly related to the causes of urban expansion, using an aggregate measure is likely to be sensitive to the scale of the study area. \citet{taubenbock2009urbanization} found urban structure to be scale dependent in an analysis of Indian cities. Six concentric rings of 10km increments were used to show heterogeneity in built-up density. They concluded that overall, the urban structure is scale dependent. These findings are consistent with previous work which showed that landscape metrics are scale dependent and the use of metric scaleograms is necessary to adequately quantify spatial heterogeneity \citep{wu2002empirical}. These issues of scale and extent relate to the issue of modifiable areal unit problem (MAUP) \citep{Openshaw1984}.

Several studies examined urbanisation utilising landscape metrics taking account of differences between the core, suburbs and periphery. \citet{schwarz2010urban, kasanko2006european} recommend delineating cities using buffers around the central business district (CBD) or examine the gradient. Improving on only reporting landscape metrics for the entire urban area, \citet{seto2005quantifying} used three buffer zones to examine landscape metrics, however the same distance buffers are used for cities of different populations. The focus of the study was the change of landscape metrics across time and cities and not whether they are scale dependent. \citet{arribas2011sprawl} made the distinction between the core and non-core across 209 European urban regions and cores using CORINE, Urban Morphological Zones (UMZ) and Urban Audit data for period 1999-2002. Multiple indexes were measured; connectivity (average commute time), decentralisation (population of people living in the non-core as a share of those living in the core), density using only urban area, scattering (ratio of patches to population), availability of open space and land use mix using Simpson's index of diversity. A self-organising map (SOM) algorithm was used to group cities in supra-national regions and observe regional patterns and overall urban sprawl. 

Studies have attempted to account for these scale issues by using a radial approach. \citet{schneider2008compact} used 1km rings to examine urban expansion, urban density, fragmentation and population density. The cut-off of the urban core was the point at which urban land density fell below 50\% . Areas outside of this core were divided into three 8 km rings representing the fringe, periphery and hinterland. The core distances for cities ranged from 3-27km. Only US and Chinese cities were found to exhibit regional differences/similarities. Urban density was found to increase the most in the core followed by the fringe, for all groups of cities. \citet{guerois2008built} measured built up land using CORINE data and 1km concentric rings. They found a steep decline in built-up area before a leveling off: two gradients were used, one for the central area (steep), and one for the periphery (shallow curve), with a first break at the historic centre of the city. The authors conclude that the geographical space is 'still shaped by the attractive power of the city centres' \citep{guerois2008built}. \citet{jiao2015urban} overcame scaling issues by using a monocentric analysis combined with a model fitting for Chinese cities. The majority of cities reduced compactness and became more dispersed over the period 1990-2010, with more dispersion occurring in the latter decade. Urban land use was found to decrease from the city centre according to the inverse s-shape rule. See table \ref{refs} in the appendix for a selection of relevant studies.

What is clear from the review of studies is that many studies have focused on landscape metrics or the change in urban land cover over time. For temporal studies, finding consistent data is a challenge, which impacts the level of data disaggregation and detail. Whereas others have examined the intra-urban structure and radial profiles, the present work is the first to examine such a large sample of cities, at a small scale (141m rings as opposed to 1km). 

In this study the heterogeneity of European cities is examined. Heterogeneity in the sense of the difference between a city's observed artificial land use (ALU) profile and the average ALU profile for 585 European cities. This measure of heterogeneity will indicate whether a city is over or under artificialised compared to the average. Points on a curve which fall above or below the average may be considered to be over or under expanded. Levels of over expansion in the suburbs and periphery are consistent with patterns of urban sprawl. 

Attempting to identify patterns in very large datasets is a difficult process. Methods such as clustering have typically been used to group observations together. Clustering algorithms are data driven, removing typology type grouping from the process. This paper introduces a conceptual framework to group cities together based upon their heterogeneity. Utilising the level of urbanisation as measured by the percentage of artificial surface within concentric circles, the difference to the average across 585 European cities is examined. This will enable cities to be classified as being more compact in the core, more sprawled in the periphery, or perhaps both. Using a scaling methodology developed by \citet{Lemoy2018} enables us to compare cities of various sizes in a consistent manner. Urban scaling helps to identify deviations by using scale independent urban indicators (SAMIs) to measure cities \citep{bettencourt2016urban}. Combining scaling with an ordering technique, has the ability to use both nomothetic and idiographic approaches to geographic information \citep{goodchild2001geographer}. Nomothetic in the sense that we are generalising cities irrespective of size, so that they can be classified, and idiographic in that specific characteristics of generalised groups or individual cities can then be analysed. 

\section{Concepts}

The homothetic scaling of artificial land use can be expressed with mathematical relations. Lemoy and Caruso (2018) found that the radial artificial land use profiles s(r) of different cities are quite similar if the distance r to the city center is rescaled to a distance r$'$ given by 
\begin{equation} \label{eq1}
r'=r \times k \textrm{ with } k=\sqrt{\frac{N_{\textrm{London}}}{N}},
\end{equation}
where $N$ is the population of the city being analysed and $N_{\textrm{London}}$ the population of London, the largest city in the dataset, which is arbitrarily chosen as a reference. $k$ is the rescaling factor: $k=1$ for London.

\begin{figure}[H]
    \includegraphics[width=\linewidth]{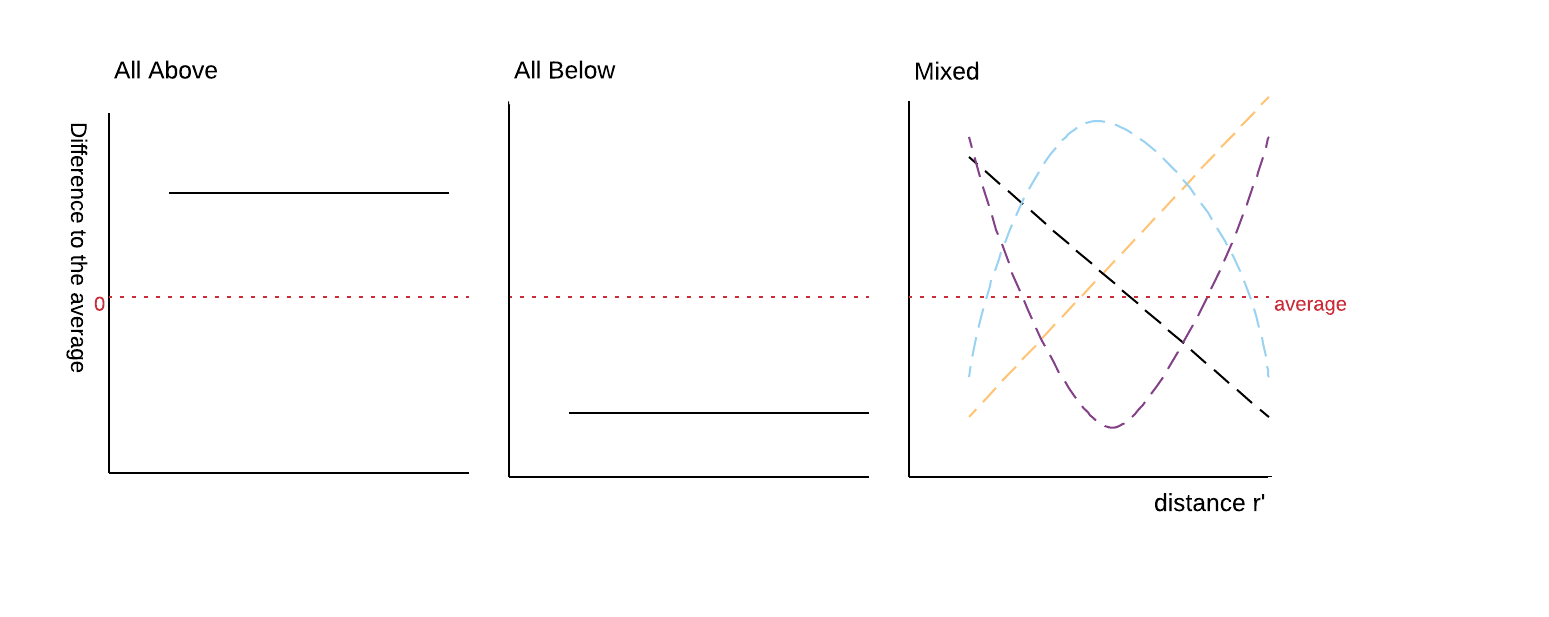}
    \caption{First tier of the two-tier profile classification}
    \label{fig:classc1}
\end{figure}

We note $s(r')$ the ALU share at rescaled distance $r'$ and $\overline{s}(r')$ the average (over all cities) ALU share at distance $r'$. We study heterogeneity using the difference $h(r')=s(r')-\overline{s}(r')$ between $s(r')$ and $\overline{s}(r')$, for each city, at each rescaled distance $r'$. This difference to the average (or heterogeneity) $h(r')$ is used in a two-tier classification, in an attempt to group similar cities. The first tier classification examines the sign and the magnitude of the heterogeneity $h(r')$. Cities are classified based on whether the difference to the average $h(r')$ is always positive (that is, positive for all values of the distance $r'$ to the CBD), always negative or a mix between the two. Figure \ref{fig:classc1} shows schematically how these differences to the average profile (represented by a red dotted line) may appear. 
For the mixed category, the heterogenity profile $h(r')$ will cross the zero horizontal axis at least once.

\begin{figure}[H]
    \includegraphics[width=\linewidth]{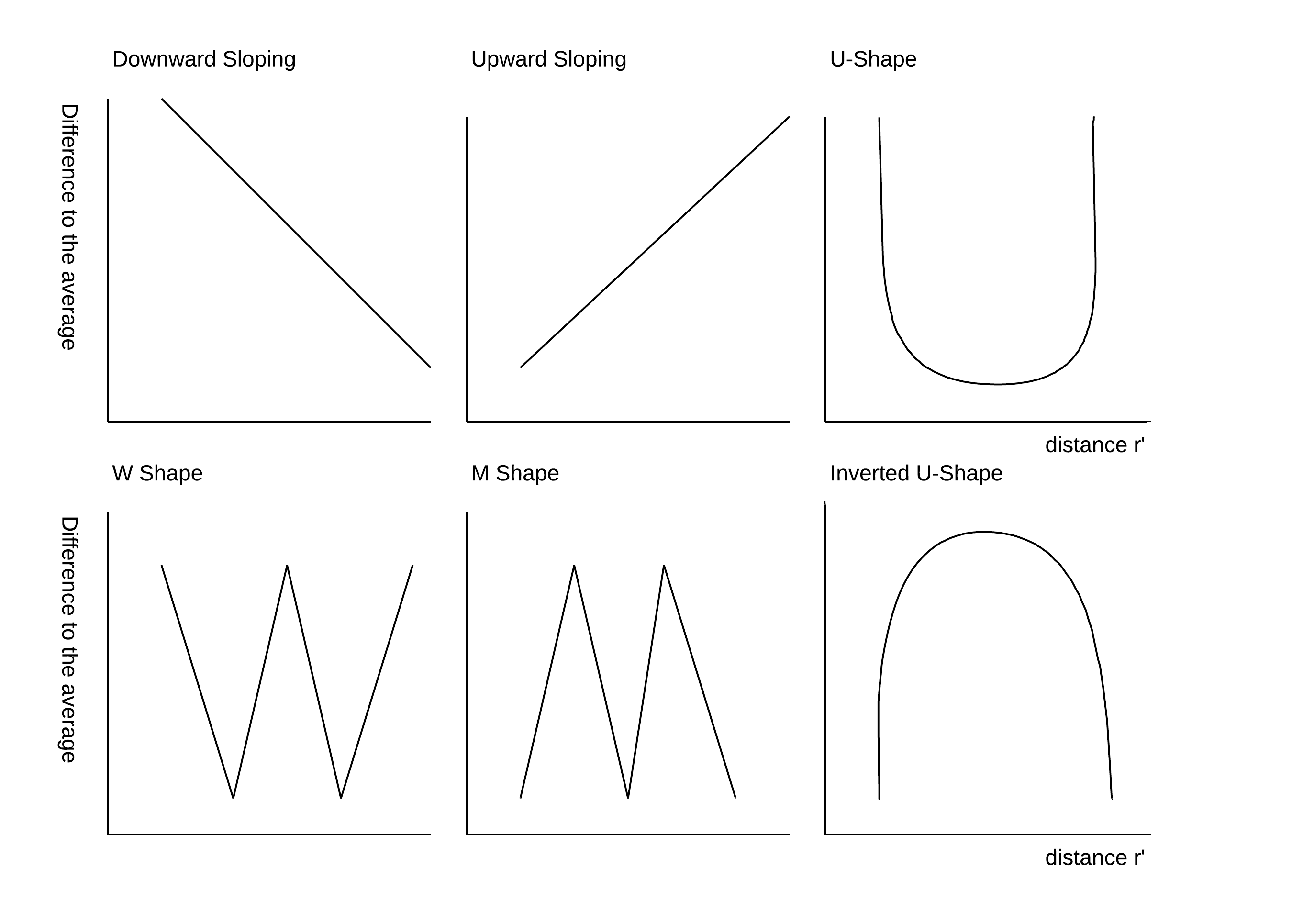}
    \caption{Second tier of the two-tier profile classification}
    \label{fig:classc2}
\end{figure}

The second tier classification examines the form of the difference to the average profile for each city. Figure \ref{fig:classc2} conceptualises the trends expected in difference to the average profiles. As the second tier classification focuses on the shape or form of the profile, the profile can be categorised as downward or upward sloping however all values can be positive or negative in relation to the average. Second tier classification describes whether the periphery, suburbs or core of a city is more or less urbanised in relation to each other. In that sense second tier classification examines more within city compared to the first tier. Colouring is added to the visualisation of the two-tier classification with red used to represent values above the average and blue to represent values below the average. Within each grouping the possibility exists of having differences in the magnitude to which a city corresponds to that classification, in which case an ordering method offers a useful method to order cities, improving our understand of the nature of ALU, urban expansion and urban sprawl and its determinants. Using a two-tier classification introduces a hierarchy of measures as a city can be both all positive/negative and also one of the six second tier classifications.

\section{Methods \& Data}

\begin{figure}[htp]
    \includegraphics[width=\linewidth]{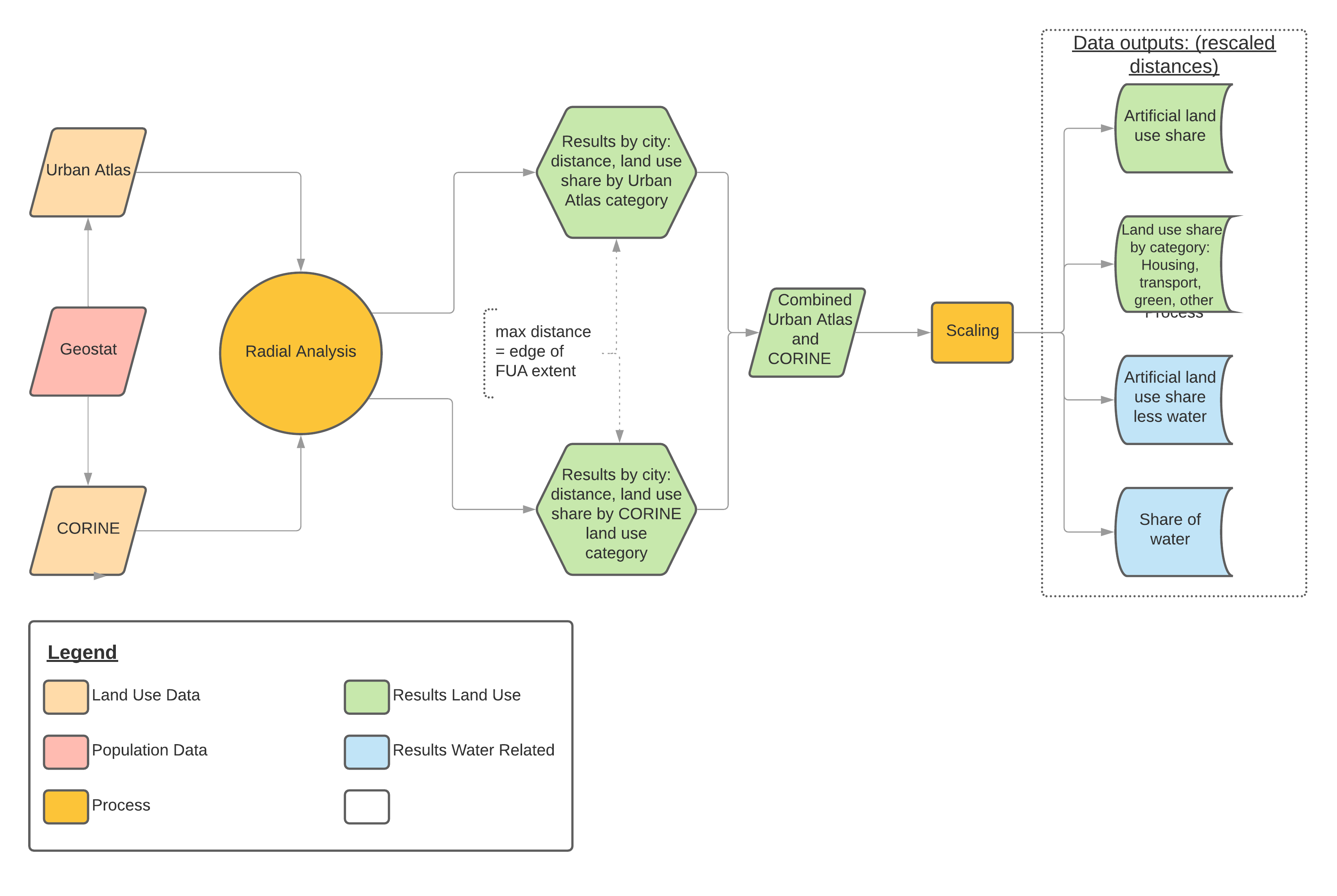}
    \caption{Pre-processing and Scaling Procedure}
    \label{fig:flow}
\end{figure}

The scaling methodology developed by \citet{Lemoy2018} and used in this analysis, requires a city's population. For this purpose the functional urban area (FUA) boundary, as defined in the urban atlas database is used to delineate urban areas. Population data from the Eurostat GEOSTAT 1km grid is down-scaled to residential land use areas from the urban atlas. A city's population is then calculated as the total down-scaled population located within a city's FUA. To overcome issues related to areas which fall outside the FUA boundary but within a concentric ring, CORINE data is combined with Urban Atlas data. With the CORINE-Urban Atlas dataset, any concentric rings drawn around the CBD will have complete land use information. This will also prove useful for defining coastal and non-coastal cities. \ref{fig:flow} details the procedure and steps involved in calculating the ALU profiles for each city. In the following sections the data used (Urban Atlas, CORINE and GEOSTAT) is discussed followed by the processing and the radial analysis. This is followed by a discussion on how the scaling law is applied. The final subsections describe the characterisation and grouping of city profiles.

Figure \ref{fig:flow} outlines the procedure used to calculate the ALU city profiles.

\subsection{Data}

The land use data for the cities examined is from the 2012 EU Copernicus Urban Atlas (UA) \citep{Copernicus2016}. A subset of cities for which population data was available for 2011 from the GEOSTAT 1km grid were used \citep{Eurostat2012}. Of the 785 FUAs available, 585 are examined which fit the criteria of being in the 2012 UA and 2011 GEOSTAT. The GEOSTAT 1B project is carried out by the European Forum for Geography and Statistics (EFGS) and Eurostat. GEOSTAT reports the number of inhabitants in a 1km grid cell for EU27 countries (plus Switzerland and Norway). Data is either aggregated (bottom-up) or disaggregated to the grid cell. The aggregation method depends upon the size of the local administrative units 2 (LAU2). The disaggregation method utilises land cover data such as the Corine or JRC Global Human Settlement Layer data to attribute inhabitants to buildings. Another hybrid method uses both aggregation and disaggregation methods with the objective of maximising data quality in GEOSTAT \citep{Eurostat2020}.

The urban atlas uses the OECD-EC definition of a FUA that takes into account density ($>$1,500 per sq km), an urban centre (with a population $>$=50,000) and also the commuting zone \citep{Dijkstra2012}. The dataset has a minimum mapping unit of 0.25 ha for urban areas and 1 ha for rural areas with exceptions for roads. In the UA land use classes are categorised into five broad themes; artificial surfaces, agricultural, natural areas, wetlands and water. From table \ref{tbl:uacodes} we focus on artificial surfaces and more specifically a range of land use classes contained within. In total 17 land use classes make up artificial surfaces, of which we focus on 12 classes where buildings are dominant, codes starting with '11' or '12'. Included in the urban fabric are both continuous urban fabric with soil sealing $>$80\% and discontinuous urban fabric 0-80\% soil sealing with a larger fraction of non-sealed or vegetation surfaces such as parks \citep{EC2016}.

Although CORINE Land Cover (CLC) has a greater spatial coverage compared to the UA, it is less precise, having a minimum mapping unit of 25 ha for areal features and 100m for linear features. CORINE has 44 land cover classes nested into three levels of aggregation/disaggregation. Table \ref{tbl:uacodes} shows the corresponding Urban Atlas classes in CORINE. There are five high level 1 classes; artificial surfaces, agricultural areas, forest and semi natural areas, wetlands and water bodies \citep{Copernicus2019}.
\clearpage
\begin{landscape}
\begin{table}
\centering
\begin{tabular}{@{}llllll@{}}
\toprule
Code & Urban Atlas Level 3 & Code & CORINE Level 3 & Land Use & Population\\&&&&&weighting \\
\cmidrule(r){1-1}\cmidrule(lr){2-2}\cmidrule(l){3-3}\cmidrule(l){4-4}\cmidrule(l){5-5}\cmidrule(l){6-6}
11100 & Continuous urban fabric (S.L. $>$ 80\%) & 111 & Continuous urban fabric & Housing & 0.85\\ 
11210 & Discontinuous dense urban fabric (S.L. 50\% - 80\%) & 112 & Discontinuous urban fabric & Housing & 0.65\\ 
  11220 & Discontinuous medium density urban fabric (S.L. 30\% - 50\%) & 112 & Discontinuous urban fabric & Housing & 0.40\\ 
  11230 & Discontinuous low density urban fabric (S.L. 10\% - 30\%) & 112 & Discontinuous urban fabric & Housing & 0.20\\ 
  11240 & Discontinuous very low density urban fabric (S.L. $<$ 10\%) & 112 & Discontinuous urban fabric & Housing & 0.05\\ 
  11300 & Isolated structures & N/A & N/A & Housing & \\ 
  12100 & Industrial, commercial, public, military and private units & 121 & Industrial or commercial units & Other & \\ 
  12210 & Fast transit roads and associated land & 122 & Road and rail networks and associated land & Transport &\\ 
  12220 & Other roads and associated land & 122 & Road and rail networks and associated land & Transport &\\ 
  12230 & Railways and associated land & 122 & Road and rail networks and associated land & Transport &\\ 
  12300 & Port areas & 123 & Port areas & Transport &\\ 
  12400 & Airports & 124 & Airports & Transport &\\ 
  13100 & Mineral extraction and dump sites & 131 \& 132 & Mineral extraction sites \& Dump sites & Other &\\ 
  13300 & Construction sites & 133 & Construction sites & Other & \\ 
  13400 & Land without current use & N/A & N/A & Other &\\ 
  14100 & Green urban areas & 141 & Green urban areas & Green &\\ 
  14200 & Sports and leisure facilities & 142 & Sport and leisure facilities & Green &\\
   \bottomrule
\end{tabular}
  \caption{Urban Atlas and Corine corresponding land use codes}\label{tbl:uacodes}
\end{table}
\end{landscape}

\subsection{CBD locations}

For the radial analysis a focal point representing the CBD is required. In order to chose a CBD representative point in a consistent manner the location of the historic city hall is used. The city hall is typically located in a highly urbanised area with high population density. Use of the city hall has been used by others  \citep{walker2018locating, wilson2012patterns}  and has been shown to be robust \citep{Lemoy2018}. In some cases due to a subsequent relocation of the city hall, it is not located in the CBD. Each CBD point is inspected using a combination of, the city land use profile curve and a visual inspection in Google Maps by uploading a KML file of the coordinates. If the point of the city hall is located outside of the CBD, a point within the historic city centre is used such as a city square or cathedral.

\subsection{Processing ALU}

The Urban Atlas vector data is converted to 20m resolution raster data improve computing efficiency. Land use information is attributed to grid cells using a majority rule. A 20m resolution ensures small objects such as roads are preserved. The population of each 1km grid cell from GEOSTAT, is then attributed to the 20m Urban Atlas grid by attributing the weightings 0.85, 0.65, 0.4, 0.2 and 0.05 to the housing categories from table \ref{tbl:uacodes}. Concentric rings of fixed width 100$\sqrt{2}$ 141m are drawn around the CBD point.

Previous analysis had the problem of no-data within the concentric rings \citep{Lemoy2018}. There was an area within the ring, for which the land use was unknown as it fell outside the FUA. Occasionally the area outside may be as a result of topography, such as water or mountains. As we use a radial monocentric analysis, the shape of the FUA is of concern. The shape may not conform to a uniform radial pattern. To overcome these issues, CORINE data for 2012 is combined with the Urban Atlas. As the UA and CORINE share classes, combining the two datasets is possible. Using the distance from the CBD to the outer most point of the FUA, all area within this ring is analysed. For areas not covered by the city UA data, CORINE data is utilised. The objective of using CORINE is to examine the impact natural features can have on the share of artificial land use. The presence of water is explicitly linked to a radial profile \citep{weiss1961distribution}. If a radial ring contains mostly water, this will give a low level of artificial land use share but does equate to an underdeveloped city. 

The level of no-data is dependent on the location of the CBD. If the CBD is close to the centroid of the FUA then the majority of no-data will be out towards the periphery. Cadiz in Spain is used to illustrate an extreme case of no-data. The CBD of Cadiz is located on a peninsula. Using a monocentric radial analysis, issues around no-data arise. By including CORINE data, the share of water within each radial ring can be quantified. The final dataset, contains 60 land use classes at $\approx$ 141m radial rings for 585 cities. By converting a 2D GIS object into a 1D curve, artificial land use as a function of the distance to the CBD can be quantified. 

\subsection{Heterogeneity Metrics}

The output from the processing and concentric rings can be thought of in terms of rings and discs. These two measures provide different insights on artificial land use: measures in discs study the share of artificial land within a given distance r from the center, while measures in rings study artificial land at a given distance r (more precisely, between r and r+$\delta$, where $\delta$ is the width of the ring). The surface of disc is given by equation \ref{eq3}

\begin{equation} \label{eq3}
V(r,i)=\pi r^2 v(r,i)
\end{equation}

where V is the surface of a disc and v is the share of the disc corresponding to land use class(es) i and radius r. In equation \ref{eq4}, the surface S of a ring can be seen as the difference between the surfaces V, of an outer disc of radius r\textsubscript{1} and an inner disc of radius r\textsubscript{2} for land use class(es) i.

\begin{equation} \label{eq4}
S(r_1,r_2,i)=V(r_2,i)-V(r_1,i)
\end{equation}

In equation \ref{eq5}, the share of a ring s for land use class(es) i, is the difference in area between two discs with radius r\textsubscript{1} and radius r\textsubscript{2} and land use class(es) i.

\begin{equation} \label{eq5}
s(r_1,r_2,i)=S(\frac{(r_1,r_2,i)}{(\pi r_2^2-\pi r_1^2)})
\end{equation}

We replace r with r’ if we are examining rescaled distances. Two metrics are used to quantify city heterogeneity. The first measure describes how much a city's ALU profile differs from the average ALU profile. Using the values for all 585, an average ALU share is calculated at $\approx$ 141m intervals up to a rescaled distance of 60km. The average is given by equation \ref{eq6}.

\begin{equation} \label{eq6}
\overline{s}(r_1,r_2,i) = \frac{\Sigma s(r_1,r_2,i)}{n}
\end{equation}

Where $\overline{s}$ is the average ALU for a specific ring and n=585, the total number of cities being analysed. Heterogeneity($\delta{s}$) is then defined as the difference in ALU for city c and the population average ($\overline{s}$) given by equation \ref{eq7}.

\begin{equation} \label{eq7}
\Delta{s}(r) = s(r_1,r_2,i,c) - \overline{s}(r_1,r_2,i)
\end{equation}

Where r is the distance examined. If $\Delta${s} is positive, the city has higher ALU than average, if $\Delta${s} is negative, the city has ALU below average. To account for water, a new ALU share is calculated. If we think about a concentric ring in a coastal area, a large share of this surface is water. If ALU share is calculated using total surface area, a low ALU share is possible, this is despite 90\% of surface land being artificialised. To overcome this issue land surface area is used to calculate a new ALU share:

\begin{equation} \label{eq8}
s^\ast(r_1,r_2,i) =  \frac{S(r_1,r_2,i)}{1 - S(r_1,r_2,w)}
\end{equation}

Where w is land use which is water and s$^\ast$ is the ALU share using only surface land area. The heterogeneity represented as difference to the average ($\Delta{s}$) is given by equation \ref{eq9}.

\begin{equation} \label{eq9}
\Delta{s}^\ast(r) = s^\ast(r_1,r_2,i,c) - \overline{s}(r_1,r_2,i)
\end{equation}

The metrics given by equations \ref{eq7} and \ref{eq9} are used to calculate heterogeneity in ALU for European cities. The results for equation \ref{eq7} are used in the two-tier profile classification.

\subsection{Two-tier profile classification}

One of the aims of this study is to attempt to group cities based on their differences to the average. Clustering methods \citep{kaufman2000finding} such as K-means \citep{forgy1965cluster}, Partitioning Around Medoids (PAM) \citep{kauf1987} and Dynamic Time Warping (DTW) \citep{keogh2005exact} are some of the possible methods and were explored. Such is the high level of heterogeneity within cities, the silhouette values calculated for K-means and PAM clustering were low ($>$0.2) indicating a weak clustering structure \citep{rousseeuw1987silhouettes}. Using DTW yielded better groupings, as evidenced in lower sum of square errors, however these are difficult to interpret conceptually. It is clear that due to the levels of heterogeneity, it is unlikely to have two cities similar across all distances to the CBD. The method of using the two-tier profile classification in combination with ordering methods is quite conceptual by nature. Such is the level of the heterogeneity between and within cities, creating such a classification using clustering methods would yield an ideal number of $\approx$ 20 groups. Some form of smoothing was therefore required in an attempt to group cities and at the same time enables the user to have a greater level of control over the method.

Using an ordering technique avoids the partitioning of cities into strict groups. To make the ordering technique more effective a two-tier profile classification is used to label cities. The first tier classification examines the sign and the magnitude of the difference in a city's ALU share to the average whereas the second tier classification examines the form of the difference to the average profile for each city. After rescaling, an ALU share at r$'$ $\approx$141m ($'$ denotes rescaled distance) intervals is available up to a r$'$=60km resulting in 601 data points for each city. Some form of smoothing is required so that cities can be categorised. Using a subset of distances 5-40km enables us to focus on the area with the highest level of heterogeneity. Four distance windows are used located at 5-14km, 14-23km, 23-31km \& 31-40km from the CBD. The median value of each window is used to represent four values for each city; a,b,c \& d. A city is categorised in the first tier: all above (all values are positive), all below (all values are negative) or mixed (a mix of positive and negative); in the second tier: 'upward sloping' a$>$b$>$c$>$d, 'downward sloping' a$>$b$>$c$>$d, 'u-shape' (a$>$b \& c$>$d), 'inverted u-shape' (a$>$b \& c$>$d) and 'w' or 'm' shape (a$>$b \& b$>$c \& c$>$d). The four points (a,b,c,d) used to create the classification located at 10,20,30 and 40 km from the CBD and can be labelled as the core (a), suburbs (b and c) and periphery (d). The two-tier profile classification uses a combination of data analysis (pre-processing, scaling and smoothing) and user interpretation (defining the different classifications).

\subsection{Orderings}

Using the groupings from the two-tier profile classification, an ordering is performed within each city grouping. The two-tier classification and subsequent ordering, is a descriptive method of ordering a list of observations, where observations are arranged in such a way that they are close to similar objects \citep{marquardt1978advances}. Unlike clustering methods, observations are displayed in a continuous manner along a line.

Unlike the two-tier profile classification which used four points, the ordering uses 35 points representing the difference to the average over distances of 5-40km with the historic centre removed (0-5km). The procedure begins with the creation of a symmetric dissimilarity matrix using euclidean distance between objects. This symmetric dissimilarity matrix is known as "two-way one-mode data since it has columns and rows (two-way) but only represents one set of objects (one-mode)" \citep{Hahsler2008}. A symmetric dissimilarity matrix uses both columns and rows but the ordering is only carried out on the rows \citep{Hahsler2008}. In the case of cities, the rows represent each city while the columns represent the level of artificial land use at that distance. The distance between each row is measured using the desired method, Euclidean, the output is the similarity of the land use profiles between those cities. We end up with a n x n matrix which is used in the ordering. The dissimilarity matrix in R, calculates the square root of the sum of the square distances between observations (cities). Using this dissimilarity matrix a rank is created for all cities within each second tier classification, with 1 representing the most similar city. Using the city with largest dissimilarity as the starting point, each subsequent city selected is the most similar to the previous city in the order. This same procedure is followed until all cities within the classification have been ordered.

\section{Results}

The results consist of three sections. (1) overall level of heterogeneity among cities (2) Differences by city grouping and country (3) Impact of water (4) Results of two-tier classification.

\subsection{Scaling Profiles}

\subsubsection{Overall}

\begin{figure}[H]
\begin{subfigure}{0.5\textwidth}
\includegraphics[width=\linewidth,height=5cm]{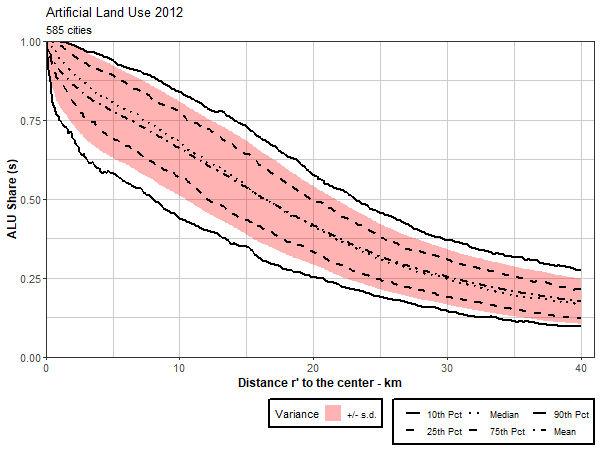}
\caption{Artificial land use share (s)}
\label{fig:percentiles}
\end{subfigure}
\begin{subfigure}{0.5\textwidth}
\includegraphics[width=\linewidth,height=5cm]{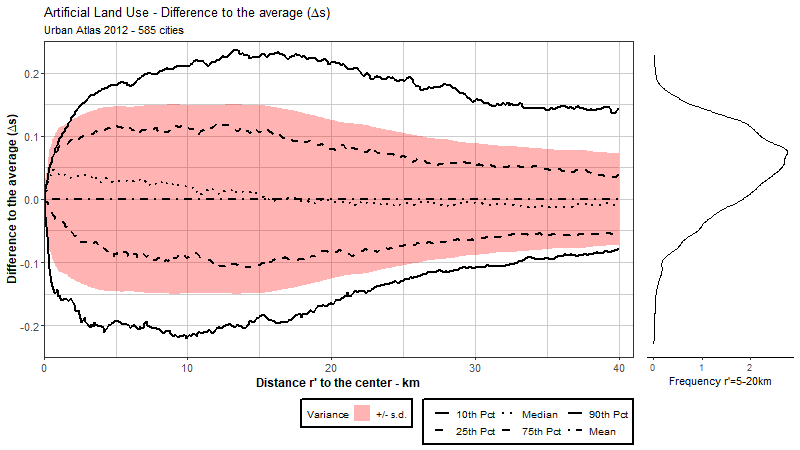}
\caption{Difference to average $\Delta{s}$}
\label{fig:quart}
\end{subfigure}
\caption{All cities - Percentiles(\nth{10},\nth{25},\nth{75},\nth{90}), mean, median and +/- standard deviation(red ribbon)}
\label{fig:pcts}
\end{figure}

We first examine the overall share of artificial land use with respect to distance to the CBD in figure \ref{fig:percentiles}. Between r$'$=5km and r$'$=15km the variation between land use profiles is at its greatest. The median value is above the mean which indicates there are a number of cities with extremely low ALU skewing the mean between r$'$=0-10km, the \nth{10} percentile also reflects this. Overall there is a steady decline in ALU with respect to the CBD, at r$'$ $\approx$25km the curve starts to decrease at a decreasing rate. The heterogeneity between cities is clearly observed, cities reach a 0.5 share of ALU between r$'$ $\approx$8km and r$'$ $\approx$23km for the top 10\% and bottom 10\% of cities.

The difference between a city's land use profile and the average land use profile is calculated for all cities. Using the 585 values at each distance a range of summary statistics are calculated and are shown in figure \ref{fig:quart}. The figure shows that the greatest differences to the average, both positive and negative, occur between the CBD and r$'$=15km. At r$'$ $\approx$10km the heterogeneity between cities is as high as 40\%. The median turns negative beyond r$'$ $\approx$15km. Focusing on the \nth{90} and \nth{10} percentile at r$'$ $>$30km the \nth{90} percentile cities are greater than the average to a larger magnitude compared with the \nth{10} percentile is less than the average. The density plot of differences to the average between r$'$=10-12km also shows a skew towards positive differences to the average. This is reflected by the median being greater than the mean between these distances.

\subsubsection{City grouping - size, country and water}

\begin{figure}[H]
    \includegraphics[width=\linewidth]{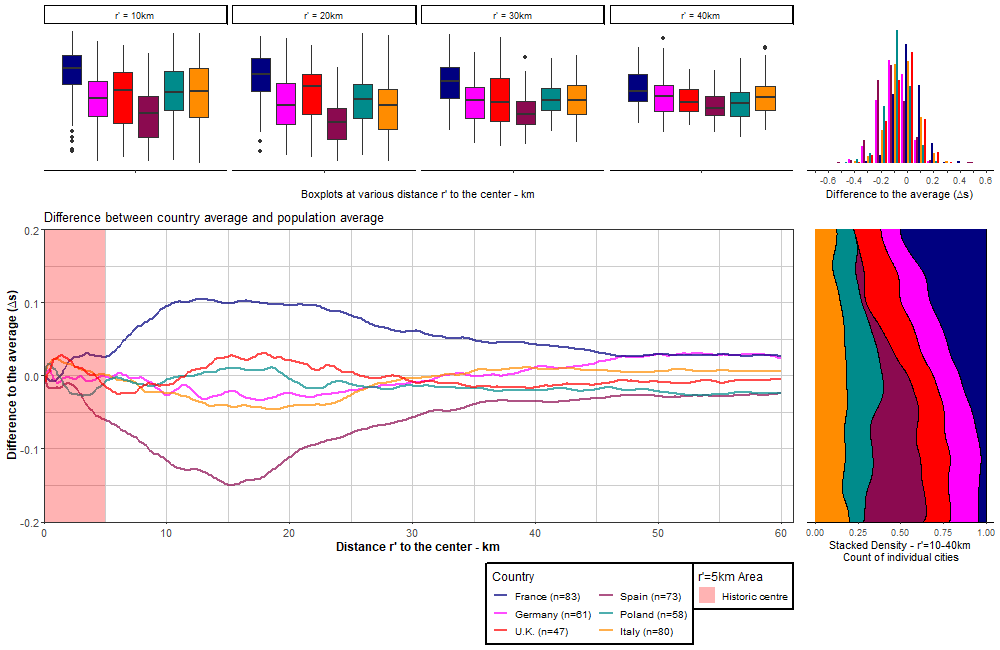}
    \caption{Heterogeneity by country}
    \label{fig:crty}
\end{figure}

The overall distribution of land use share and difference to the average has shown the levels of heterogeneity that exist between cities. Using groupings we attempt to understand these differences. Does the heterogeneity of a city have similarities with another city which it shares a particular aspect with. For example, do cities have similar differences based upon location or country, city size grouping, perhaps a historical grouping or if they have similar physical conditions. Figures \ref{fig:sizec}, \ref{fig:crty}, \ref{fig:ctrwat} and \ref{fig:profclass} all have the same layout. The main plot (bottom left) shows the curves of differences to the average, the top left shows boxplots of the differences at specific distances, the top right is a histogram showing the frequency of the values and the bottom right figure shows a stacked density plot highlighting the share of differences. All distances in the figures are rescaled distances.

\subsubsection{City size}

\begin{figure}[H]
    \includegraphics[width=\linewidth]{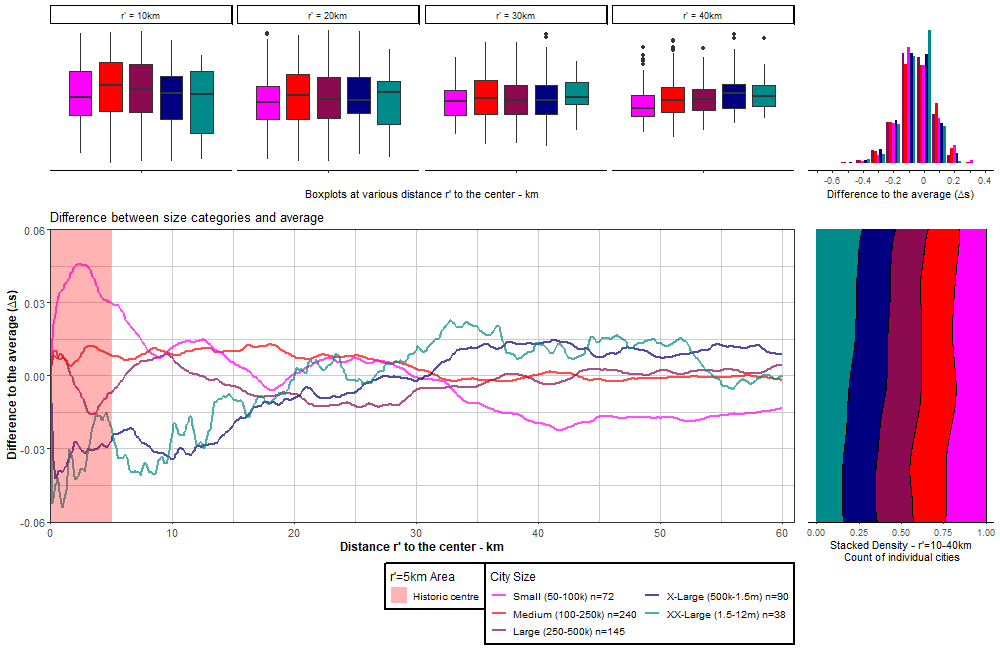}
    \caption{Heterogeneity by city size grouping}
    \label{fig:sizec}
\end{figure}

The first grouping examined is that of city size. From figure \ref{fig:sizec}, small cities are more artificialised around the CBD and 'historic centre' (r$'$ $>$ 5km) compared to other city size categories, however there is a lot of variation as noted by the height of the boxplots. Beyond r$'$=30km there is a sorting of city size from largest to smallest, with the largest cities have an ALU share above average and the smallest cities being the least artificialised beyond this distance. From r$'$ 30-60km small cities are less artificialised on average. From the stacked density chart, we can see that Xx-large cities are more artificalised than the average. The opposite is the case for small cities. Overall, the differences between city size categories is small, as the median of each category falls within one standard deviation of every other group (boxplots). Within each size category there are cities which are more or less artificialised compared to the average. The scaling law appears to control for city size effectively. Given city size is relatively poor at explaining ALU heterogeneity, other groupings are explored.

\subsubsection{Country}

Unlike the city size categories, grouping cities by country highlights clear differences in ALU heterogeneity among cities. We focus on the six countries with the highest share of cities, these countries account for 69\% of the 585 cities in the dataset. Across all distances r$'$=5km to r$'$=50km, France and Spain are the two countries most dissimilar to each other. French cities are the most artificialised on average across all distances, reaching a maximum of 0.1 difference in ALU share to the average. The opposite is the case for Spanish cities, at r$'$=15km Spanish cities have 0.15 less ALU share compared to the average (Fig \ref{tbl:aerialmed} in appendix shows aerial photos of different rescaled distances). Deducing from this information, at r$'$=15km, French cities have on average an ALU share that is 0.25 higher than that for Spanish cities. German and Italian cities are both below the average at r$'$ $\approx$  15km, whereas at the same distance UK cities are above the average. From the boxplots at distances r$'$=10km and r$'$=20km, there is clearly high within country variation among UK cities with a skew towards high levels of ALU share. The stacked density chart shows that Italy has the most even spread of cities that are above and below the average. Spanish cities are more frequently below average, whereas French cities are more frequently above the average.

\subsubsection{Impact of water}

We attempt to understand the differences between countries by exploring one possible root cause, the share of water within each concentric ring. Utilising the combined CORINE-Urban Atlas data it is possible to calculate the share of water within each ring across all distances. The question of whether coastal cities are less urbanised than non-coastal cities is examined.

As coastal cities do not have a full disc but a segment, they artificialise more land further away from the CBD to have the same level of urbanisation as a non-coastal city. Given a disc contains more space compared to a segment, coastal cities compensate for having less space by building further away from the centre compared to if the city had a complete disc.
For a specific quantity of a disc, to fit the same area on a segment, you would have to increase the radius, hence you are urbanising further away from the centre. scaling allows us to compare cities in a consistent matter, this allows us to examine urbanisation between countries. 
A hybrid dataset which combines Urban Atlas and CORINE data was used to examine ALU before and after accounting for the level of water. From this dataset we were able to determine the share of water within a concentric ring. As the level of water within a ring varies across space, the share of water within a disc is used to calculate the share of water. Taking a r$'$=20km, the share of water for each city is calculated. Cities are divided into four groups using Jenks cut-offs of 0.0, 0.13, 0.33 and 0.78. We label these Water 1, 2, 3 and 4, with 4 being the group with the highest share of water.

\begin{figure}[H]
\begin{subfigure}{\linewidth}
\includegraphics[width=\linewidth]{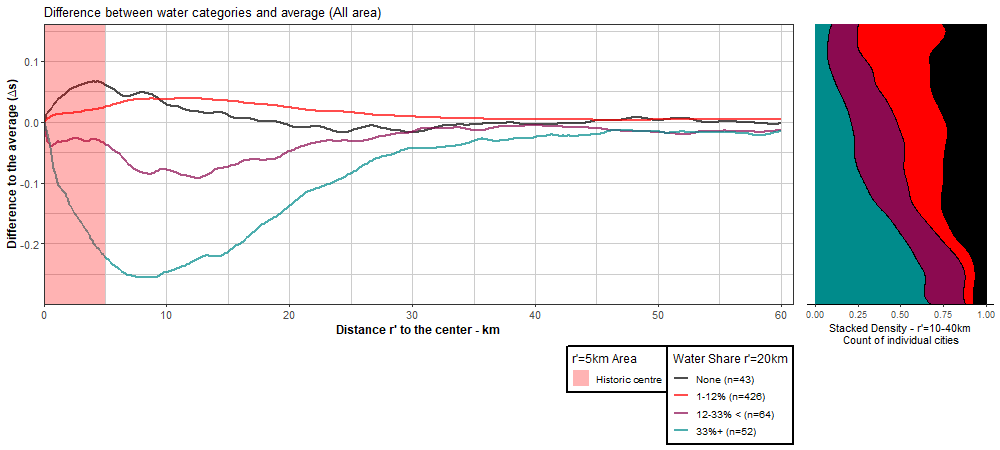} 
\caption{All area considered}
\label{fig:subim1wat}
\end{subfigure}
\\ 
\begin{subfigure}{\linewidth}
\includegraphics[width=\linewidth,height=5cm]{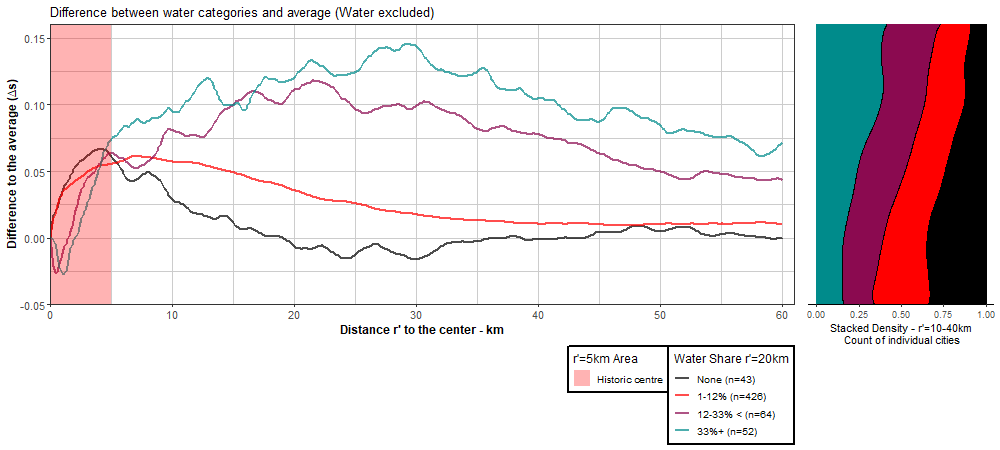}
\caption{Only land area}
\label{fig:subim2wat}
\end{subfigure}
\caption{Difference to the average ($\Delta{s}$) by water category - Before and after accounting for water}
\label{fig:watg}
\end{figure}

From figure \ref{fig:watg} we can see that after grouping cities, the share of water has a determining impact on the difference in ALU to the average. From figure \ref{fig:subim1wat}, cities with the highest share of water have the lowest levels of ALU. Both groups of cities with greater than 12\% water have below average ALU beyond r$'$=15km. There is a clear sorting by the level of water, except in the case of 'none' and between 1-12\% water, the latter category includes cities which have a river and are not 'coastal' by definition. For these cities the level of water does not appear to have an affect. From the density plot, we are able to examine share of cities by difference to the average. At -0.2 to the average, almost half of these cities are in the 33\% + water group. Towards the upper end with over 0.1 difference to the average, cities with 12\% and less make up the majority.

In figure \ref{fig:subim2wat} the level of water is removed from the calculation of the ALU share. After removing water the group with $>$33\% has the highest levels of ALU above average. This trend appears to suggest that coastal cities of water are more compact compared to non-coastal cities or indeed cities with just a river. At r$'$=25km the 33\% + group has an ALU share that is $\approx$0.12 greater than the 'none' grouping. When using just land area, the groupings with the highest share of water (12-33\% and 33\% +), have the highest count of cities with an ALU share $>$0.10 (stacked density, figure \ref{fig:subim2wat}).

\begin{figure}[H]
    \includegraphics[width=\linewidth]{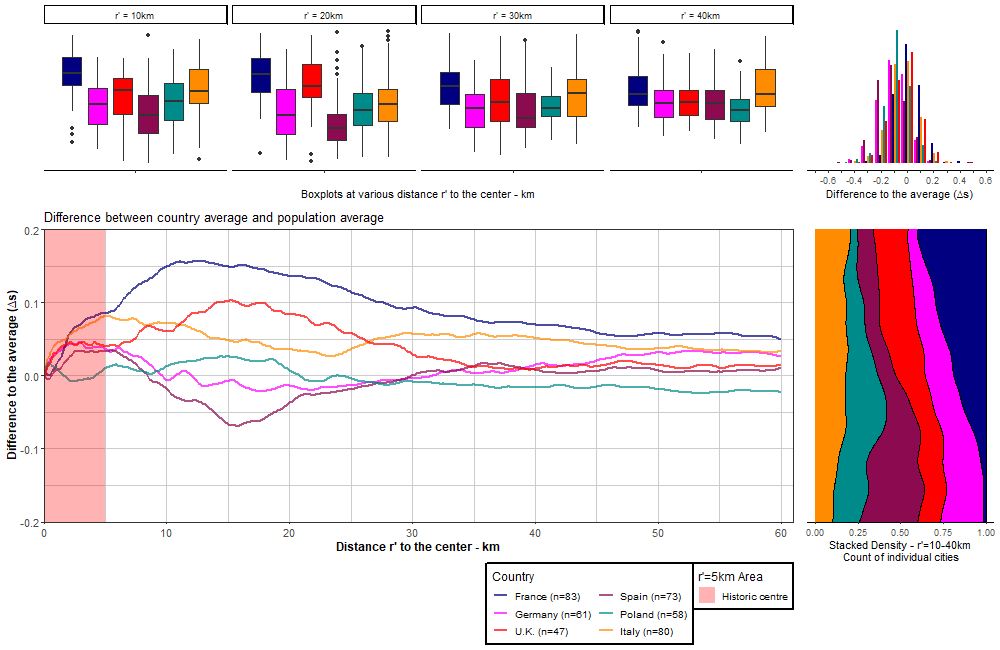}
    \caption{Heterogeneity by country - land area only (water not included)}
    \label{fig:ctrwat}
\end{figure}

We re-examine figure \ref{fig:ctrwat} by removing the share of water to explain the differences between French and Spanish cities. From figure \ref{fig:ctrwat} there is a clear impact of water on average levels of ALU by country. This change is greater for the minimum Spanish ALU share (-.15 to -.07) compared to the maximum French ALU share (.11-.16). ALU share for Spain however remains largely negative. With the exception of Germany and Poland, the UK, France and Italy all have ALU shares above the average. Italy has experienced the biggest change, at r$'$=15km from -0.05 to 0.05. From boxplots r$'$=20km, the UK no longer has an upward skew as shown by the boxplots. At the upper end of the stacked density plot, $>$0.1, French cities hold the majority. UK cities have increased their share at the expense of Italian, Polish and Spanish cities. Polish and German cities are more frequently around 0.0 - -0.1, this is due to a low number of coastal cities in these countries
. 

\subsection{Two-tier profile classification}

In this section city profiles are grouped according to classification identified in figures \ref{fig:classc1} and \ref{fig:classc2}, using the sign (first tier) and shape (second tier) of the differences to the average curve. Table \ref{tbl:countclass} shows the count of cities by profile classification for the first and second tier. For first tier classification there are approximately as many cities all above the average, as there are all below. The majority of cities cross the average curve and are a mix of both positive and negative differences to the average. For the second tier, the majority of cities are either a u-shape or inverted u-shape. A u-shape is above average in the core, below average in the suburbs and above average in the periphery. The opposite is the case of the inverted u-shape. Table \ref{tbl:countclass} highlights the level of heterogeneity both within and between cities.

\begin{table}[H]
\centering
\begin{tabular}{@{}llll@{}}
\toprule
First Tier & Count of cities & Second Tier & Count of cities\\
\cmidrule(r){1-1}\cmidrule(lr){2-2}\cmidrule(l){3-3}\cmidrule(lr){4-4}
All above & 147 & Downward sloping & 90\\
All below & 157 & Upward sloping & 97\\
Mixed & 281 & Inverted u-shape & 161\\
 & & U-shape & 204\\
 & & W or m shape & 33\\
Total & 585 & Total & 585\\
   \bottomrule
\end{tabular}
  \caption{Count of cities by Two-tier profile classification}\label{tbl:countclass}
\end{table}

\begin{figure}[H]
\begin{subfigure}{\linewidth}
\includegraphics[width=\linewidth]{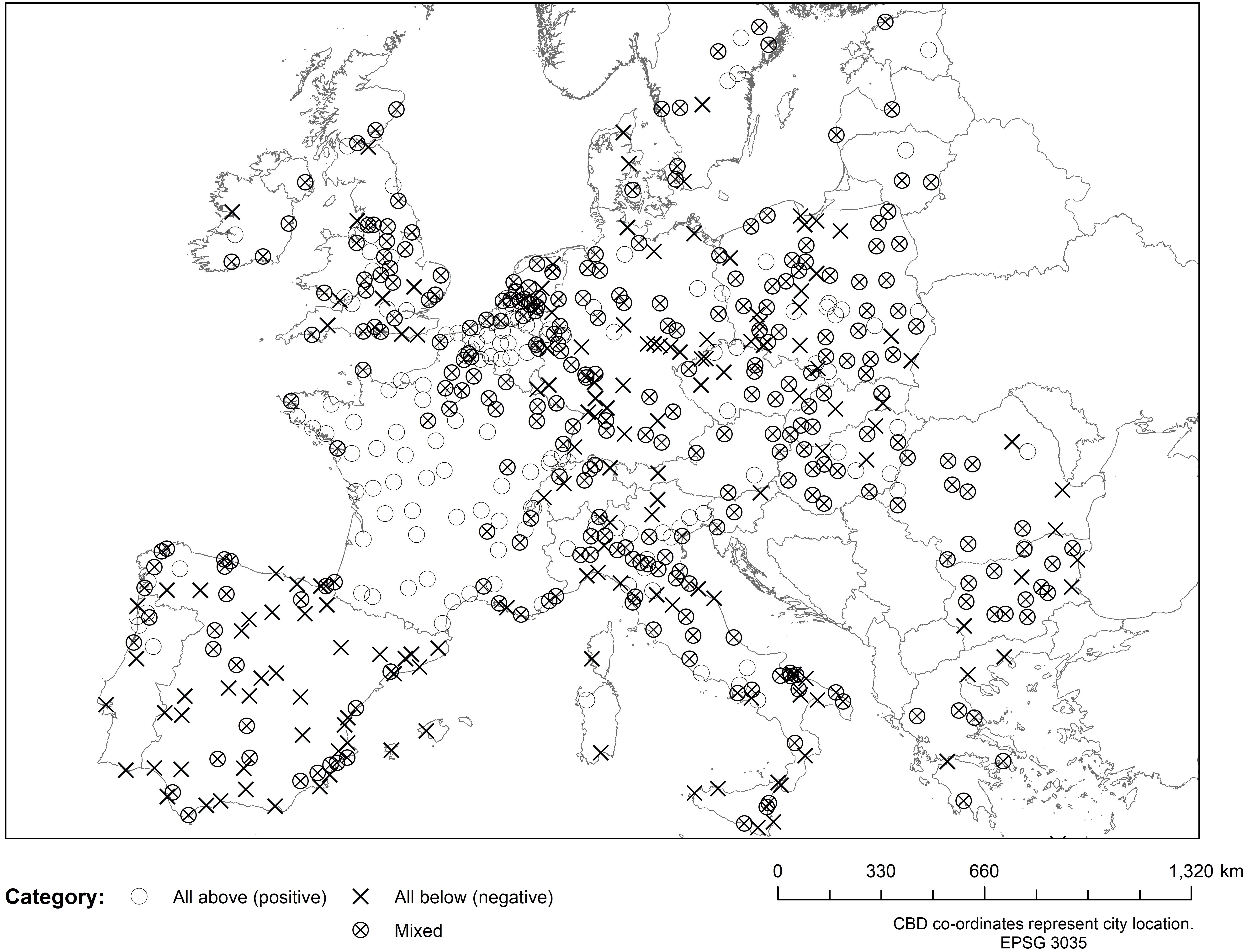} 
\caption{First tier classification}
\label{fig:ftc-city}
\end{subfigure}
\\ 
\begin{subfigure}{\linewidth}
\includegraphics[width=\linewidth]{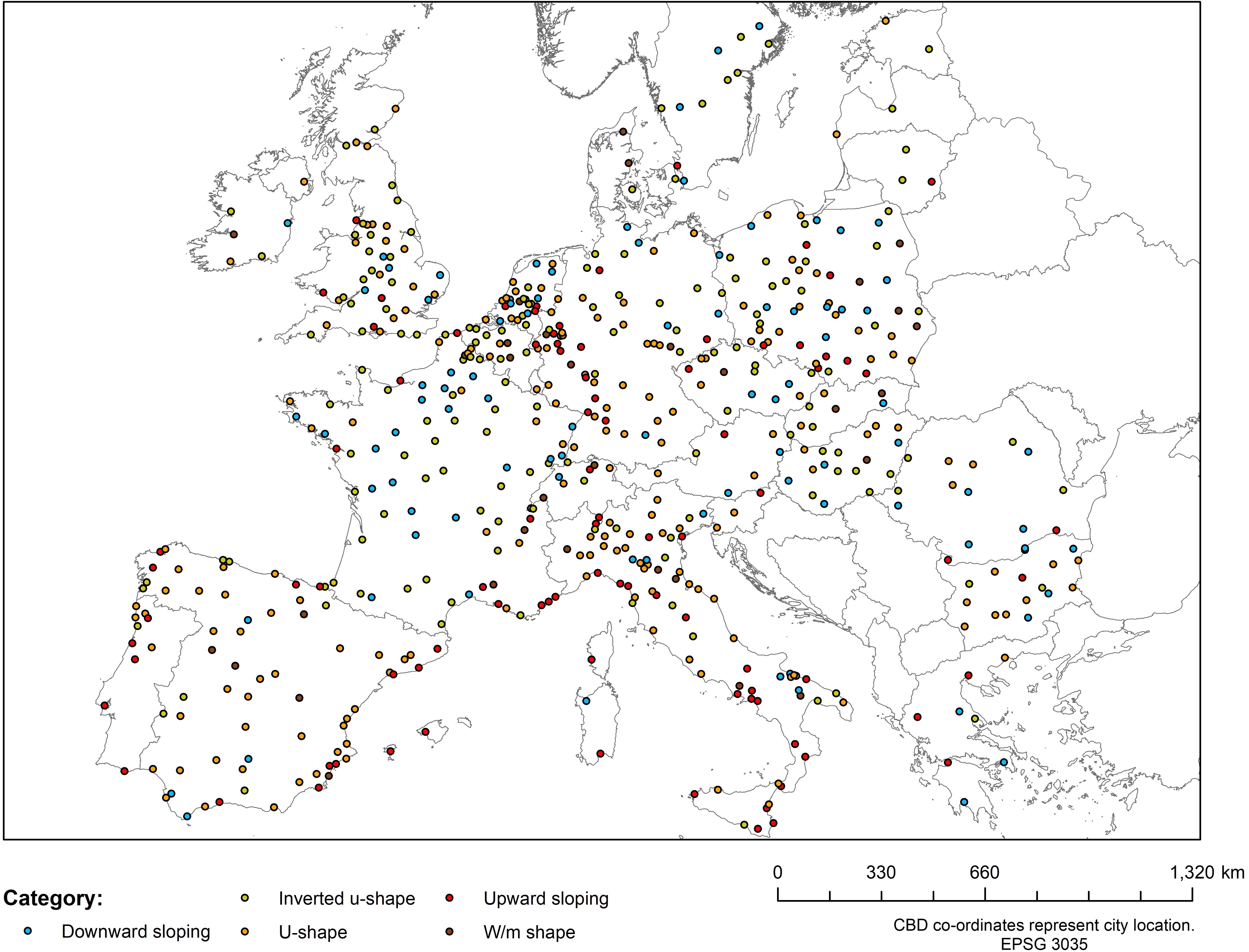}
\caption{Second tier classification}
\label{fig:stc-city}
\end{subfigure}
\caption{Spatial distribution of cities by two-tier profile classification}
\label{fig:classmap}
\end{figure}

Figure \ref{fig:classmap} shows the spatial distribution of cities by two-tier profile classification. The map highlights the heterogeneity that cities between cities. One notable aspect is the high number of first tier 'all above' cities in France and the high number of 'all below' cities in Spain (Figure \ref{fig:ftc-city}). This is consistent with our earlier findings that French cities are more urbanised than Spanish cities. Outside of these two countries the pattern becomes difficult to decipher. In relation to figure \ref{fig:stc-city}, French cities are predominantly 'downward sloping' and 'inverted u-shape', whereas Spanish cities are 'u-shape' and 'upward sloping'. We cannot say with certainty that a country or geographical region is of a particular second tier classification. The high levels of heterogeneity, justifies further the use of a ordering technique to order cities within classification to aid with understanding similarity of cities.

\clearpage

\begin{figure}[H]
    \includegraphics[width=\linewidth]{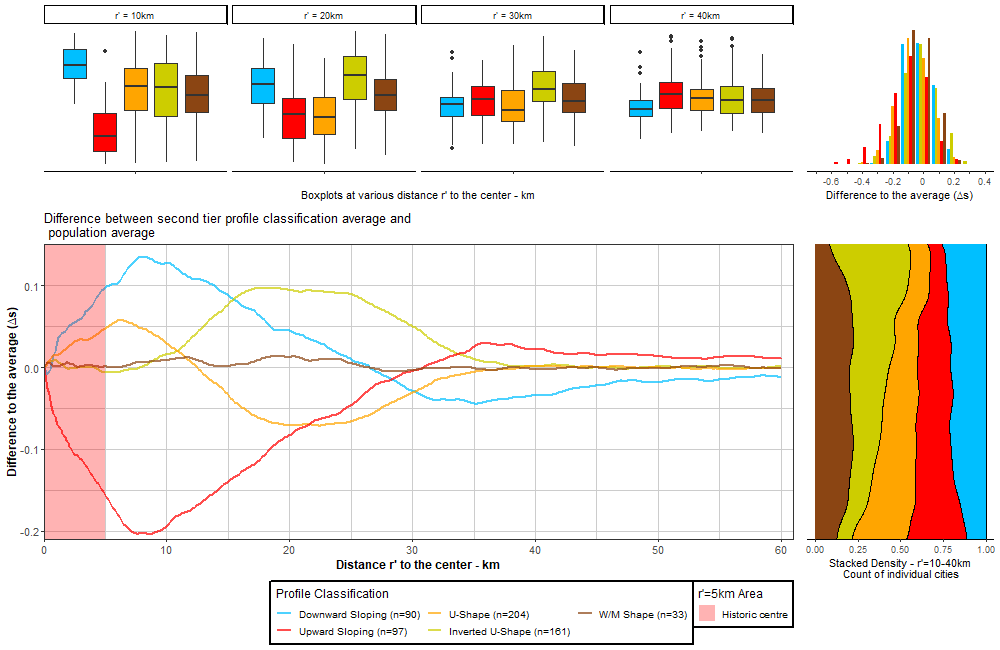}
    \caption{Difference between second tier profile classification group average and population average}
    \label{fig:profclass}
\end{figure}

Figure \ref{fig:profclass} details summary statistics of the different second tier classifications. Boxplots at r$'$=10k show that 'downward sloping' has the highest difference to the average values and 'upward sloping' the lowest. From the profiles, at distance r$'$ $\approx$20km the 'downward sloping' and 'inverted u-shape' curves cross. The maximum difference between the 'inverted u-shape' and 'u-shape' is reached at r$'$=22km, a difference of $\approx$0.18. The profiles shapes, fit the shape of the curves the classification represents; i.e. downward sloping follows a downward trend and u-shape is in the shape of a 'u'. The stacked density plot details which cities appear most frequently at various values. At 0.1 difference to the average, 'inverted u-shape' and 'downward sloping' cities are the most frequent cities. At -0.1 difference to the average 'u-shape' and 'upward sloping' cities are the most frequent. 'Downward sloping' is the most desirable form, as it is compact at the core and has less over-urbanisation at the periphery. 'U-shape' is also desirable as it has a compact core and less sprawl in the suburbs compared to the 'inverted u-shape'. 'Upward sloping' is a undesirable form, the core is under urbanised on average with the suburbs and periphery at or above the average. Due to the 'random' nature of the 'w/m shape', it is difficult to draw conclusions. Overall the form seems to follow the average ALU for European cities.

\subsubsection{Ordering}

Using the second tier profile classification of cities, an ordering method is used to order each within classification list. Differences to the average are colour coded, with 'red' representing above average and 'blue' representing below average. Colour gradients are scaled using +/- four standard deviations from the mean value of zero. Table \ref{fig:profclass} shows the five profile classifications using a user ordering method. The vertical axis details the city's name, while the horizontal axis shows the rescaled distance from r$'$=0-40km (A detailed list of cities along with their difference to the average values a,b,c,d and first and second profile classification, is given in the appendix). 
For 'downward' and 'upward', we can see the transition from red to blue and blue to red respectively. 'Inverted u-shape' starts off blue before transitioning to red and then back to blue again, in other cases the profile is predominantly red. There are some blue/red pixels which are smoothed out when the profile is reduced to four points. As we are ordering using a large number of points (77), they become visible. The majority of 'u-shape' cities transition from red to blue and back to red or light blue. Table \ref{tbl:long} in the appendix shows a complete list of all cities and their corresponding first and second tier classification.

\begin{landscape}
\begin{table}[h!]
  \centering
  \begin{tabular}{ | c | c | c | c | c |}
    \hline
    Upward & Downward & Inverted u-shape & U-shape & W/M shape \\ (n=82) & (n=58) & (n=166) & (n=151) & (n=128) \\ \hline
    \begin{minipage}{.2\textwidth}
      \includegraphics[width=\linewidth, height=90mm]{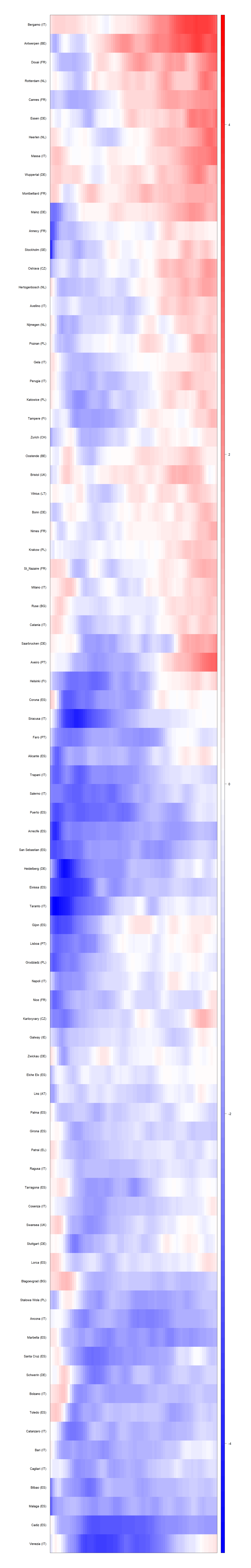}
    \end{minipage}
    & 
    \begin{minipage}{.2\textwidth}
      \includegraphics[width=\linewidth, height=90mm]{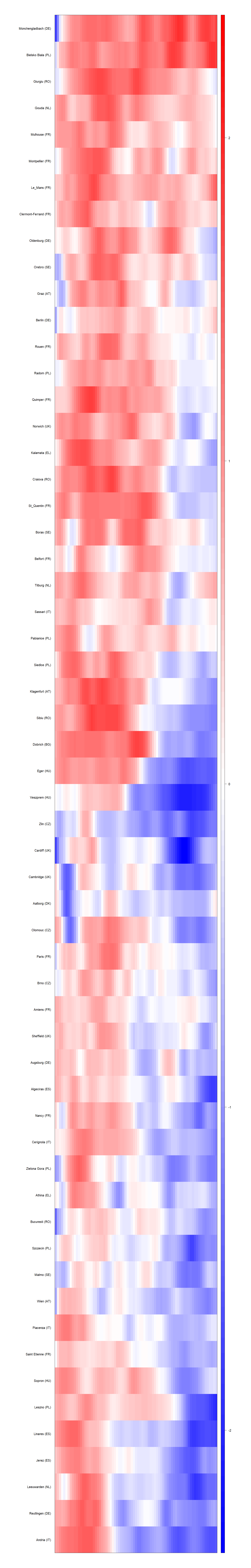}
    \end{minipage}
    & 
    \begin{minipage}{.2\textwidth}
      \includegraphics[width=\linewidth, height=90mm]{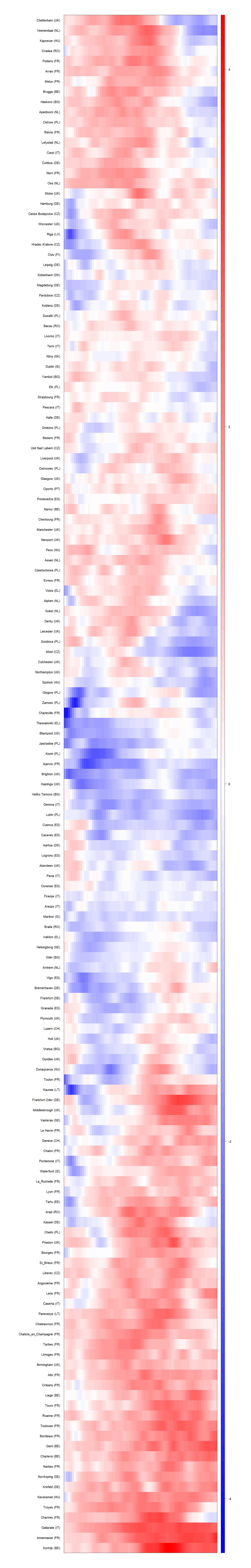}
    \end{minipage}
    & 
    \begin{minipage}{.2\textwidth}
      \includegraphics[width=\linewidth, height=90mm]{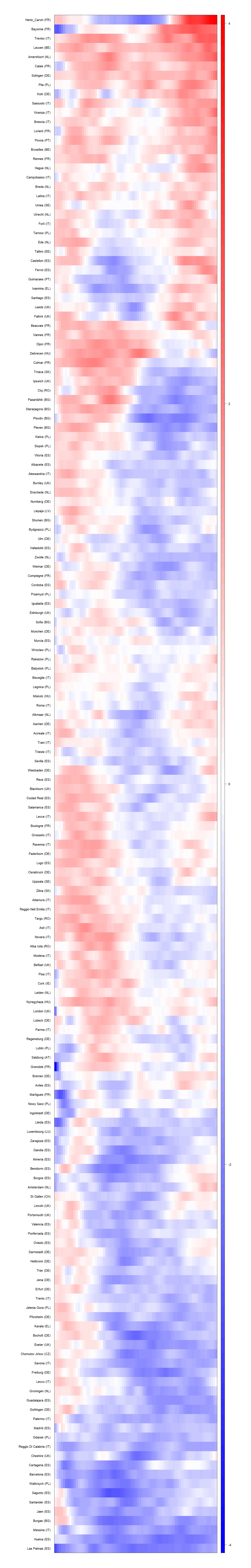}
    \end{minipage}
    & 
    \begin{minipage}{.2\textwidth}
      \includegraphics[width=\linewidth, height=90mm]{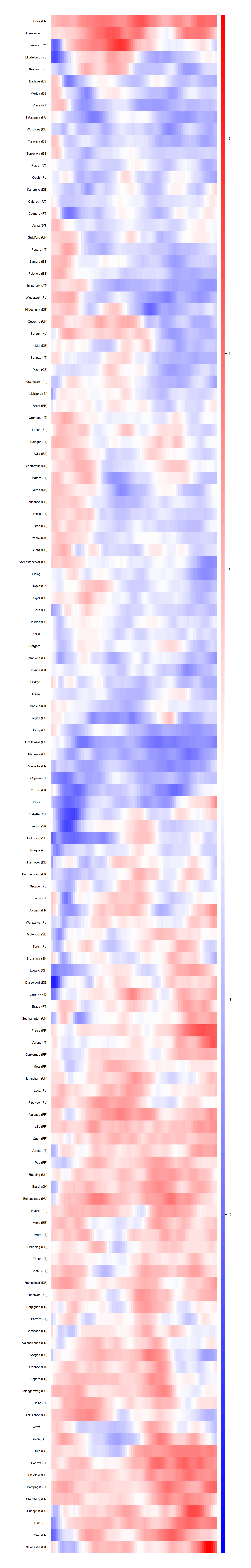}
    \end{minipage}
    \\ \hline
  \end{tabular}
  \caption{Ordering City List - Second Tier Categories}\label{tbl:myLboro}
\end{table}
\end{landscape}

\section{Conclusion}

Using a radial analysis of cities, this paper has examined the difference in artificial land use share to the average for 585 European cities. A scaling methodology was adopted to control for city size, placing the focus on the magnitude of the difference to the average and the shape of the land use profiles. 

Our results are particularly useful to urban economists, policymakers and geographers as cities are bench marked based upon their difference to the average profile and also the shape of the land use profile. It presents a useful starting point to policymakers and planners as to which cities in Europe are most 'similar' to theirs. Previously, they may have used city size which we have shown is not a useful predictor of similarity. Examining the within city structure is more useful.This enabled us to show clear differences between countries and coastal and non-coastal cities. The differences between (under urbanised) and France (over urbanised) warrants further examination, possibly through the use of additional socio-economic and demographic variables. The second tier classification based on urban form, enabled us to group cities with similar internal ALU patterns. Within each classification the ordering method aided in understanding the differences between cities.

\clearpage
\section{References}
\bibliography{hetero}

\section{Abbreviations}

ALU Artificial Land Use
CBD Central business district
LUZ Large urban zone
FUA Functional urban area
UA Urban atlas
UMZ Urban Morphological Zones

\section{Appendix}

\begin{landscape}
\begin{table}[ht]
\begin{tabular}{|p{3cm}|p{2cm}|p{2cm}|p{1cm}|p{3cm}|p{4cm}|p{4cm}|}
  \hline
 Study & Data & Region & No. Cities & City boundary & Divisions & Indicators \\ 
\hline
  \citet{angel2011dimensions} & MODIS 500m & World & 3646 & urban cluster & 1 & change \\ 
  \hline
  \citet{jiao2015urban} & Landsat TM/ETM+ & China & 28 & physical and functional & 1 km rings & urban land density \\
  \hline
  \citet{schneider2008compact} & Landsat & World & 25 & functional & varying core buffer + 4, 8km rings & urban indicators \\ \hline
  \citet{wolff2018compact} & CORINE & Europe & 5692 urban areas & morphological and density & 1 & residential density \\ 
  \hline
  \citet{siedentop2012sprawls} & CORINE & Europe & - & 20km grid for the country & - & land consumption, sealing degree \\ 
  \hline
  \citet{oueslati2015determinants} & CORINE & Europe & 237 & Large urban zone (LUZ) & 1 & change and fragmentation \\ 
  \hline
  \citet{guerois2008built} & CORINE & Europe & 54 & 40-100km radius & 1 km rings & built-up area \\
\hline
\end{tabular}
\label{refs}
\caption{A selection of studies relevant to the present work.}
\end{table}
\end{landscape}

\begin{table}[H]
  \centering
  \begin{tabular}{ | c | c | c | }
    \hline
    All Artificial Land Uses & Housing & Transport \\ \hline
    \begin{minipage}{.3\textwidth}
      \includegraphics[width=\linewidth, height=40mm]{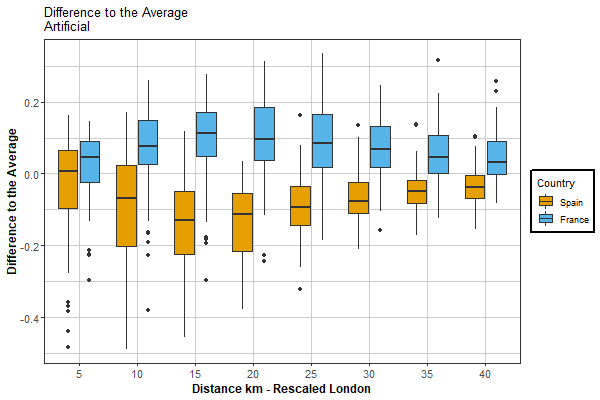}
    \end{minipage}
    &
      \begin{minipage}{.3\textwidth}
      \includegraphics[width=\linewidth, height=40mm]{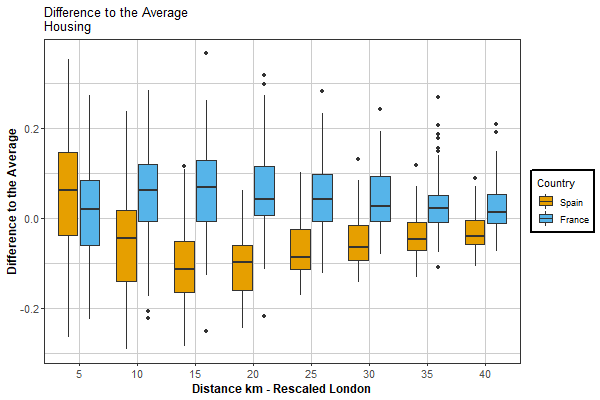}
    \end{minipage}
    & 
      \begin{minipage}{.3\textwidth}
      \includegraphics[width=\linewidth, height=40mm]{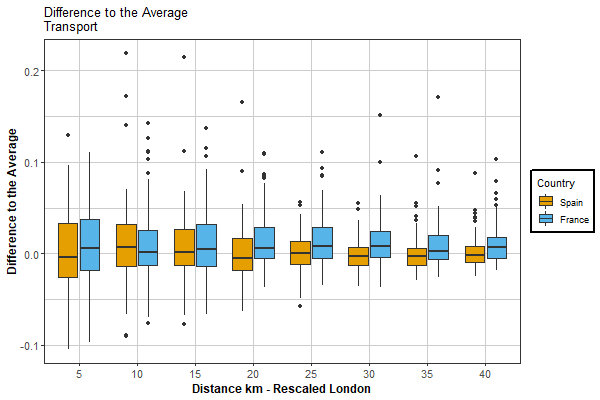}
    \end{minipage}
    \end{tabular}
    \begin{tabular}{ | c | c | c | }
    \\hline
    Other & Green space & Water \\ \hline
    \begin{minipage}{.3\textwidth}
      \includegraphics[width=\linewidth, height=40mm]{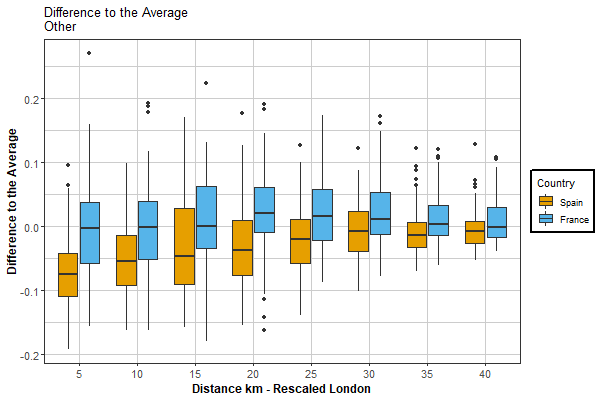}
      \\hline
    \end{minipage}
    &
      \begin{minipage}{.3\textwidth}
      \includegraphics[width=\linewidth, height=40mm]{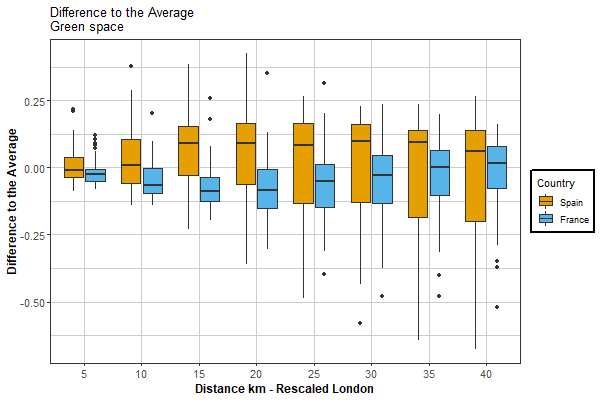}
      \\hline
    \end{minipage}
    & 
      \begin{minipage}{.3\textwidth}
      \includegraphics[width=\linewidth, height=40mm]{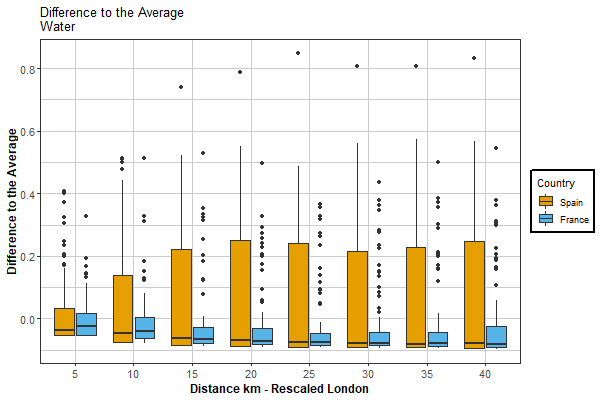}
      \\hline
    \end{minipage}
  \end{tabular}
  \caption{Country Heterogeneity by land use type}\label{tbl:myLboro}
\end{table}

\begin{table}[ht]
  \centering
  \begin{tabular}{ | c | c | c | }
    \hline
    All cities & Small & Medium \\ \hline
    \begin{minipage}{.3\textwidth}
      \includegraphics[width=\linewidth, height=40mm]{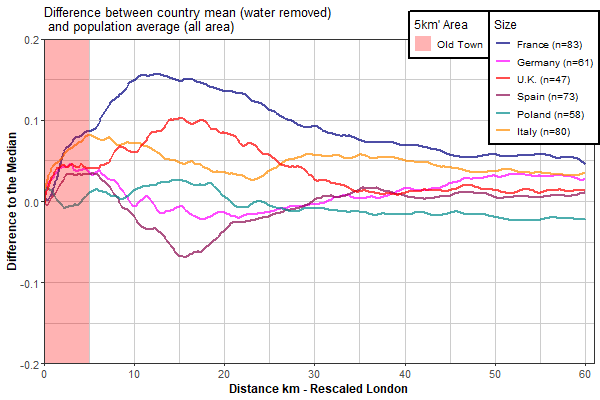}
    \end{minipage}
    &
      \begin{minipage}{.3\textwidth}
      \includegraphics[width=\linewidth, height=40mm]{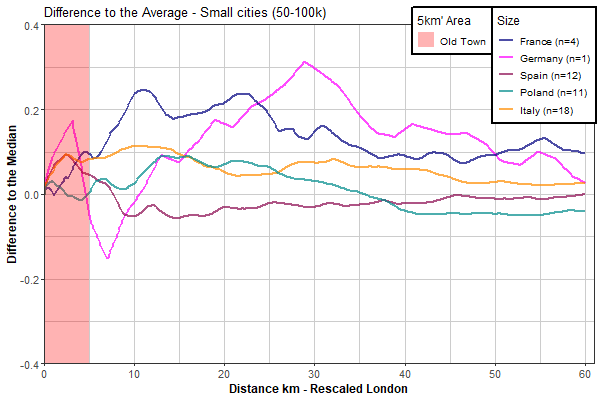}
    \end{minipage}
    & 
      \begin{minipage}{.3\textwidth}
      \includegraphics[width=\linewidth, height=40mm]{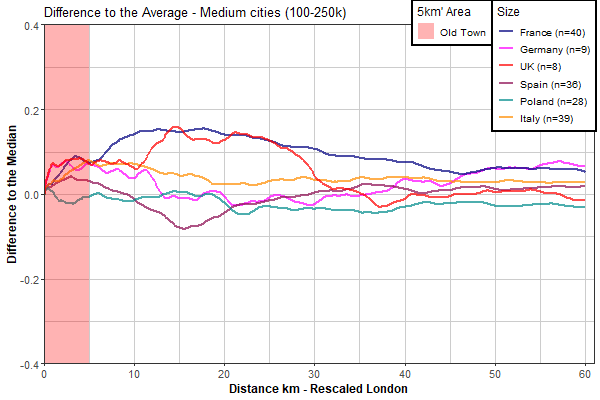}
    \end{minipage}
    \end{tabular}
    \begin{tabular}{ | c | c | c | }
    \hline
    Large & X-Large & XX-Large \\ \hline
    \begin{minipage}{.3\textwidth}
      \includegraphics[width=\linewidth, height=40mm]{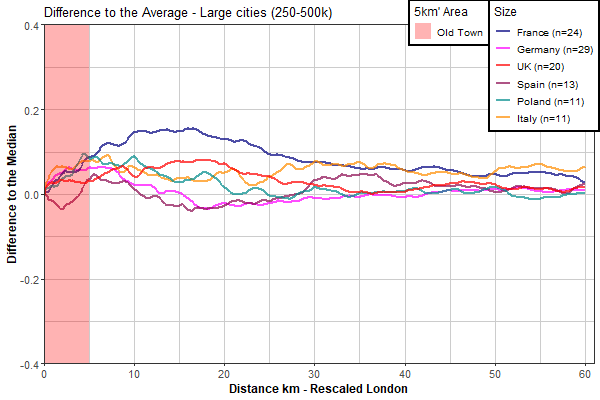}
      \\hline
    \end{minipage}
    &
      \begin{minipage}{.3\textwidth}
      \includegraphics[width=\linewidth, height=40mm]{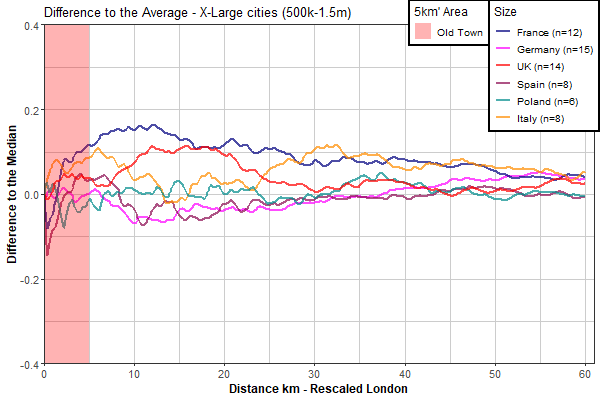}
      \\hline
    \end{minipage}
    & 
      \begin{minipage}{.3\textwidth}
      \includegraphics[width=\linewidth, height=40mm]{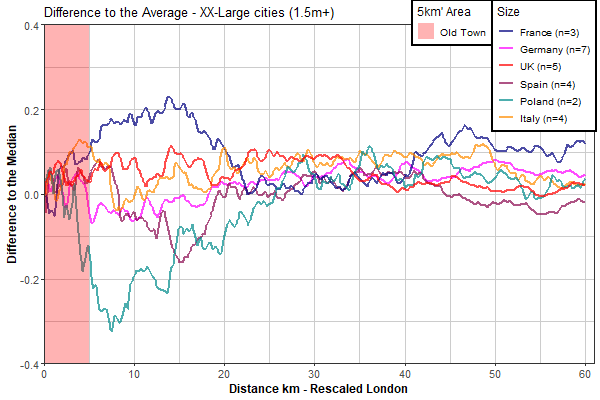}
      \\hline
    \end{minipage}
  \end{tabular}
  \caption{Country Heterogeneity by city size}\label{tbl:myLboro}
\end{table}

\begin{figure}[H]
    \centering
    \includegraphics[width=12cm, height=190mm]{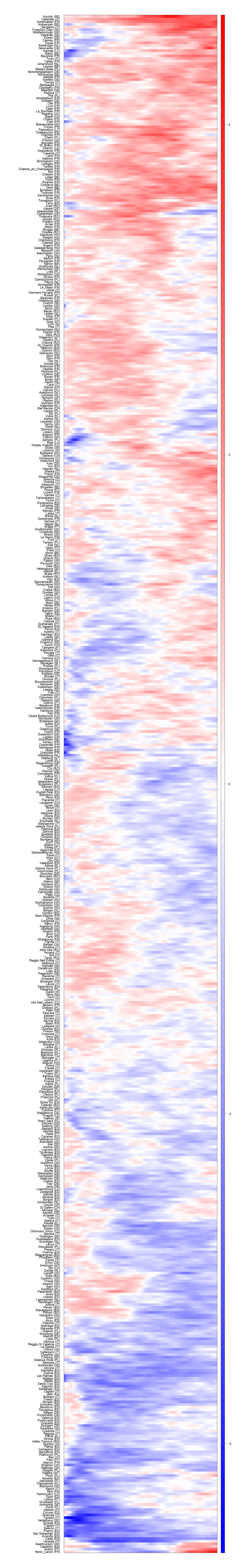}
    \caption{Ordering city list - All cities}
    \label{fig:galaxy}
\end{figure}

\begin{landscape}
\begin{table}[h!]
  \centering
  \begin{tabular}{ | c | c | c | }
    \\hline
    All above (positive) & All below (negative) & Mixed  \\ \hline
    \begin{minipage}{.2\textwidth}
      \includegraphics[width=\linewidth, height=90mm]{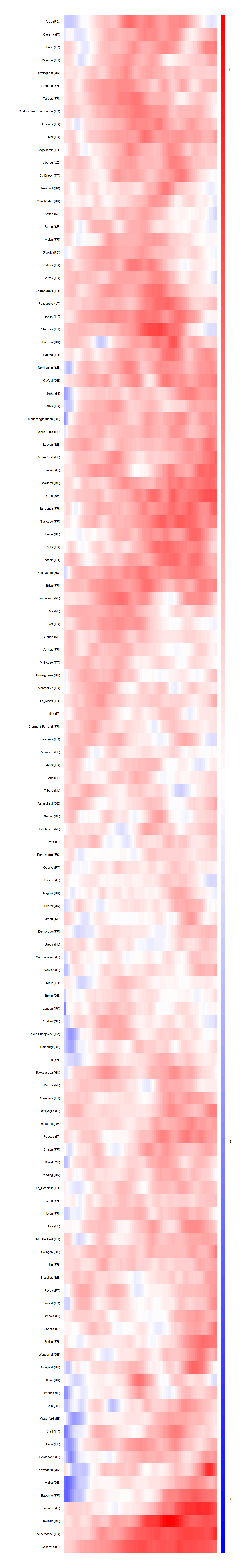}
    \end{minipage}
    & 
    \begin{minipage}{.2\textwidth}
      \includegraphics[width=\linewidth, height=90mm]{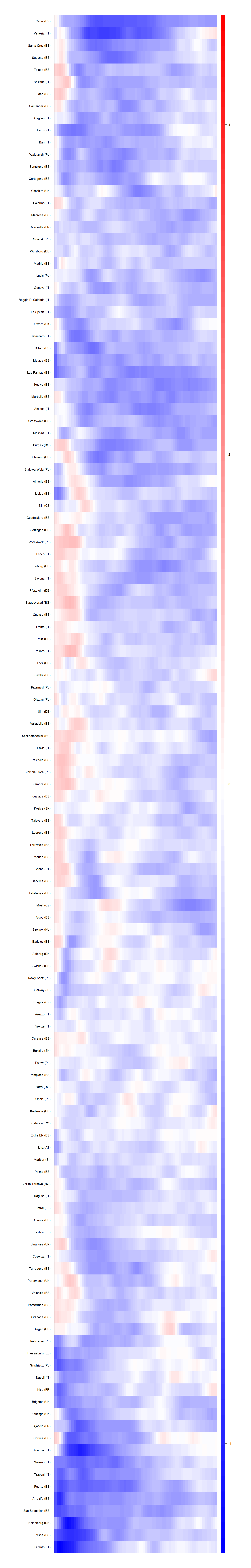}
    \end{minipage}
    & 
    \begin{minipage}{.2\textwidth}
      \includegraphics[width=\linewidth, height=90mm]{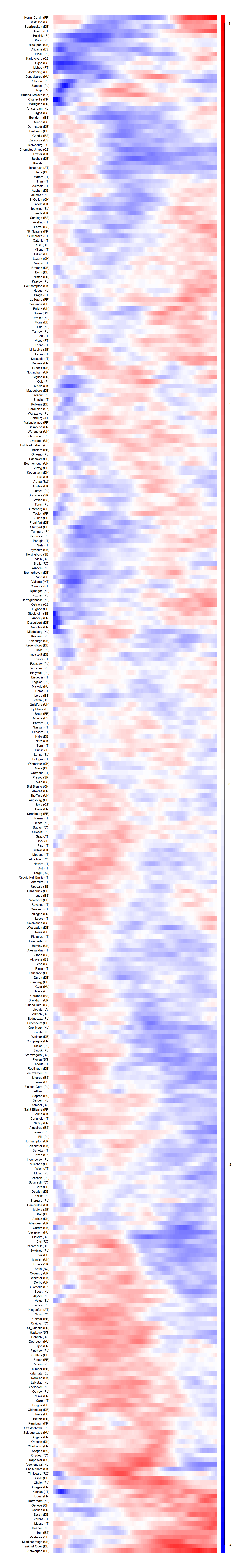}
    \end{minipage}
    \\ \hline
  \end{tabular}
  \caption{Ordering City List - First Tier Categories}\label{tbl:myLboro}
\end{table}
\end{landscape}

%
\begin{longtable}{l|rrrr|rrll}
\caption{Attributes of 585 cities. Including four difference to the average rolling mean values, population, two-tier classifications and rescaled r$'$=20km distance\label{tbl:long}}\\
  \hline
  & \multicolumn{4}{|c|}{Heterogeneity rolling mean} & & & \\
 City & 10km & 20km & 30km & 40km & r$'$20km & Pop. & \nth{1} tier cat. & \nth{2} tier cat. \\ 
  \hline
    \endfirsthead
    \hline
 City & 10km & 20km & 30km & 40km & r$'$=20km & Pop. & \nth{1} tier cat. & \nth{2} tier cat. \\ 
  \hline
  \endhead
Aachen (DE) & -0.01 & -0.05 & 0.07 & 0.14 & 4.30 & 0.53 & u & mix \\ 
  Aalborg (DK) & 0.00 & -0.04 & -0.03 & -0.03 & 3.97 & 0.45 & w/m & mix \\ 
  Aarhus (DK) & -0.08 & -0.02 & -0.03 & -0.02 & 4.77 & 0.65 & w/m & neg \\ 
  Aberdeen (UK) & -0.03 & -0.04 & -0.08 & -0.05 & 3.90 & 0.44 & u & neg \\ 
  Acireale (IT) & -0.06 & -0.06 & 0.03 & 0.07 & 1.48 & 0.06 & u & mix \\ 
  Ajaccio (FR) & -0.32 & -0.08 & -0.08 & 0.00 & 1.78 & 0.09 & w/m & mix \\ 
  Alba Iulia (RO) & 0.14 & -0.04 & -0.02 & -0.07 & 1.80 & 0.09 & w/m & mix \\ 
  Albacete (ES) & -0.03 & -0.07 & -0.02 & -0.03 & 2.58 & 0.19 & w/m & neg \\
\hline
\end{longtable}

\end{document}